\documentclass{emulateapj}

\usepackage{graphicx}
\usepackage[usenames]{color}
\usepackage{amsmath}
\usepackage{natbib}

\defcitealias{quiet:2011}{QUIET\,2011}
\defcitealias{Instrumentpaper:2012}{QUIET\,2012}

\renewcommand{\arraystretch}{1.1}
\renewcommand{\phm}{\phantom{-}}
\renewcommand{\phn}{\phantom{1}}
\newcommand{\phe}{\phantom{${}^1$}}

\begin{document}

\title{
Second Season QUIET Observations:\\
Measurements of the CMB Polarization Power Spectrum at 95\,GHz}

\author{
QUIET Collaboration ---
D.~Araujo\altaffilmark{1},
C.~Bischoff\altaffilmark{2,3},
A.~Brizius\altaffilmark{2,4},
I.~Buder\altaffilmark{2,3},
Y.~Chinone\altaffilmark{5,6},
K.~Cleary\altaffilmark{7},
R.~N.~Dumoulin\altaffilmark{1},
A.~Kusaka\altaffilmark{2,8},
R.~Monsalve\altaffilmark{9,10},
S.~K.~N\ae ss\altaffilmark{11},
L.~B.~Newburgh\altaffilmark{1,8},
R.~Reeves\altaffilmark{7},
I.~K.~Wehus\altaffilmark{12,13},
J.~T.~L.~Zwart\altaffilmark{1,14},
L.~Bronfman\altaffilmark{15},
R.~Bustos\altaffilmark{9,15,16},
S.~E.~Church\altaffilmark{17},
C.~Dickinson\altaffilmark{18},
H.~K.~Eriksen\altaffilmark{11,19},
T.~Gaier\altaffilmark{20},
J.~O.~Gundersen\altaffilmark{9},
M.~Hasegawa\altaffilmark{5},
M.~Hazumi\altaffilmark{5},
K.~M.~Huffenberger\altaffilmark{9},
K.~Ishidoshiro\altaffilmark{5},
M.~E.~Jones\altaffilmark{12},
P.~Kangaslahti\altaffilmark{20},
D.~J.~Kapner\altaffilmark{2,21},
D.~Kubik\altaffilmark{22},
C.~R.~Lawrence\altaffilmark{20},
M.~Limon\altaffilmark{1},
J.~J.~McMahon\altaffilmark{23},
A.~D.~Miller\altaffilmark{1},
M.~Nagai\altaffilmark{5},
H.~Nguyen\altaffilmark{22},
G.~Nixon\altaffilmark{8,24},
T.~J.~Pearson\altaffilmark{7},
L.~Piccirillo\altaffilmark{18},
S.~J.~E.~Radford\altaffilmark{7},
A.~C.~S.~Readhead\altaffilmark{7},
J.~L.~Richards\altaffilmark{7},
D.~Samtleben\altaffilmark{4,25},
M.~Seiffert\altaffilmark{20},
M.~C.~Shepherd\altaffilmark{7},
K.~M.~Smith\altaffilmark{2,8},
S.~T.~Staggs\altaffilmark{8},
O.~Tajima\altaffilmark{2,5},
K.~L.~Thompson\altaffilmark{17},
K.~Vanderlinde\altaffilmark{2,26},
R.~Williamson\altaffilmark{1,2}
}

\vspace{+0.2in}

\altaffiltext{1}{Department of Physics and Columbia Astrophysics Laboratory, Columbia University, New York, NY 10027, USA}
\altaffiltext{2}{Kavli Institute for Cosmological Physics, Department of Physics, Enrico Fermi Institute, The University of Chicago, Chicago, IL 60637, USA; send correspondence to I.~Buder, \textbf{ibuder@uchicago.edu}}
\altaffiltext{3}{Harvard-Smithsonian Center for Astrophysics, 60 Garden Street MS 42, Cambridge, MA 02138, USA}
\altaffiltext{4}{Max-Planck-Institut f\"ur Radioastronomie, Auf dem H\"ugel 69, 53121 Bonn, Germany}
\altaffiltext{5}{High Energy Accelerator Research Organization (KEK), 1-1 Oho, Tsukuba, Ibaraki 305-0801, Japan}
\altaffiltext{6}{Astronomical Institute, Graduate School of Science, Tohoku University, Aramaki, Aoba, Sendai 980-8578, Japan}
\altaffiltext{7}{Cahill Center for Astronomy and Astrophysics, California Institute of Technology, 1200 E. California Blvd M/C 249-17, Pasadena, CA 91125, USA}
\altaffiltext{8}{Joseph Henry Laboratories of Physics, Jadwin Hall, Princeton University, Princeton, NJ 08544, USA}
\altaffiltext{9}{Department of Physics, University of Miami, 1320 Campo Sano Drive, Coral Gables, FL 33146, USA}
\altaffiltext{10}{School of Earth and Space Exploration, Arizona State University, 781 E. Terrace Road, Tempe, AZ 85287, USA}
\altaffiltext{11}{Institute of Theoretical Astrophysics, University of Oslo, P.O. Box 1029 Blindern, N-0315 Oslo, Norway}
\altaffiltext{12}{Department of Astrophysics, University of Oxford, Keble Road, Oxford OX1 3RH, UK}
\altaffiltext{13}{Department of Physics, University of Oslo, P.O. Box 1048 Blindern, N-0316 Oslo, Norway}
\altaffiltext{14}{Physics Department, University of the Western Cape, Private Bag X17, Bellville 7535, South Africa}
\altaffiltext{15}{Departamento de Astronom\'ia, Universidad de Chile, Casilla 36-D, Santiago, Chile}
\altaffiltext{16}{Departamento de Astronom\'ia, Universidad de Concepci\'on, Casilla 160-C, Concepci\'on, Chile}
\altaffiltext{17}{Kavli Institute for Particle Astrophysics and Cosmology and Department of Physics, Stanford University, Varian Physics Building, 382 Via Pueblo Mall, Stanford, CA 94305, USA}
\altaffiltext{18}{Jodrell Bank Centre for Astrophysics, Alan Turing Building, School of Physics and Astronomy, The University of Manchester, Oxford Road, Manchester M13 9PL, UK}
\altaffiltext{19}{Centre of Mathematics for Applications, University of Oslo, P.O. Box 1053 Blindern, N-0316 Oslo, Norway}
\altaffiltext{20}{Jet Propulsion Laboratory, California Institute of Technology, 4800 Oak Grove Drive, Pasadena, CA, USA 91109}
\altaffiltext{21}{Micro Encoder Inc., Kirkland, WA 98034, USA}
\altaffiltext{22}{Fermi National Accelerator Laboratory, Batavia, IL 60510, USA}
\altaffiltext{23}{Department of Physics, University of Michigan, 450 Church Street, Ann Arbor, MI 48109, USA}
\altaffiltext{24}{Tradeworx, Inc., 10 Broad Street, Third Floor, Red Bank, NJ 07701, USA}
\altaffiltext{25}{Nikhef, Science Park, Amsterdam, The Netherlands}
\altaffiltext{26}{Department of Physics, McGill University, 3600 Rue University, Montreal, Quebec H3A 2T8, Canada}

\slugcomment{
Published as ApJ 760, 145---This paper should be cited as ``QUIET Collaboration et al. (2012)''
}
\journalinfo{Published as ApJ 760, 145---Draft version \today}

\begin{abstract}
The Q/U Imaging ExperimenT (QUIET) has observed the cosmic
microwave background (CMB) at 43 and 95\,GHz. The 43-GHz results have
been published in \cite{quiet:2011}, and here we report the
measurement of CMB polarization power spectra using the 95-GHz data.
This data set comprises 5337 hours of observations recorded by an
array of 84 polarized coherent receivers with a total array
sensitivity of 87\,$\mu$K$\sqrt{\text{s}}$. Four low-foreground fields
were observed, covering a total of $\sim$ 1000 square degrees with an
effective angular resolution of $12\farcm8$, allowing for
constraints on primordial gravitational waves and
high--signal-to-noise measurements of the $E$-modes across three acoustic peaks.
The data reduction was performed using two
independent analysis pipelines, one based on a pseudo-$C_\ell$ (PCL)
cross-correlation approach, and the other on a maximum-likelihood (ML)
approach. All data selection criteria and filters were modified
until a predefined set of null tests had been satisfied before inspecting any
non-null power spectrum.
The results derived by the two pipelines are in good agreement.
We characterize
the $EE$, $EB$ and $BB$ power spectra between $\ell= 25$ and 975 and find that the $EE$ spectrum is consistent with $\Lambda$CDM,
while the $BB$ power spectrum is consistent with zero.
Based on these measurements, we constrain the tensor-to-scalar ratio
to $r = 1.1^{+0.9}_{-0.8}$ ($r < 2.8$ at 95\% C.L.) as derived by the
ML pipeline, and $r =1.2^{+0.9}_{-0.8}$ ($r < 2.7$ at 95\% C.L.) as
derived by the PCL pipeline.
In one of the fields, we find a correlation with the dust component of
 the Planck Sky Model, though the corresponding excess power is small
 compared to statistical errors.
Finally, we derive limits on all known
systematic errors, and demonstrate that these
correspond to a tensor-to-scalar ratio smaller than $r = 0.01$, the
lowest level yet reported in the
literature. 
\end{abstract}

\keywords{cosmic background radiation---Cosmology: observations---Gravitational waves---inflation---Polarization}

\maketitle

\tighten

{\renewcommand{\thefootnote}{\fnsymbol{footnote}}}
\setcounter{footnote}{0}

 \section{Introduction}
The theory of inflation explains several well-observed properties of the 
Universe~\citep[e.g.][and references therein]{liddle:2000}: the lack of
spatial curvature, the 
absence of relic monopoles from a grand unified theory's broken 
symmetry, the large-scale correlations that imply a much larger particle 
horizon than the Big Bang scenario provides without inflation, and the 
nearly--scale-invariant Gaussian fluctuations. Although inflation was 
developed to explain these known properties of the Universe, which are 
now probed with high precision by recent cosmological
observations~\cite[]{Komatsu:2010fb,2011ApJ...739...52D,2011ApJ...743...28K,2012arXiv1203.6594A,2009ApJ...700.1097H,2009ApJS..185...32K,2010ApJ...708..645R},
the model also has a new feature: the early, 
exponential expansion of space generates a stochastic background of 
gravitational waves. In the near term, polarization measurements of the 
cosmic microwave background (CMB) present the most promising approach to 
detect these gravitational waves, which cause an odd-parity ($B$-mode)
polarization pattern on angular scales larger than a
degree~\citep[]{Seljak:1997,Kamionkowski:1996zd}. Detection (or 
non-detection) of these patterns will place strong constraints on the 
inflation paradigm.

In the slow-roll approximation~\cite[for a review see][]{liddle:2000}, the $B$-mode intensity is parametrized
by the tensor-to-scalar ratio $r$, which is related to the energy
scale $V$ of inflation by $V \sim (r/0.01)^{1/4} \times
10^{16}$\,GeV. For many classes of inflationary models, $r$ can be as
large as $0.01 \lesssim r \lesssim 0.1$~\citep[]{Boyle:2005ug}.

A combination of CMB-temperature-anisotropy measurements, baryon-acoustic-oscillation data, and supernova observations has given the
most stringent limit to date, $r \lesssim 0.2$ at 95\% confidence level
(C.L.), nearly limited by cosmic
variance~\citep{Komatsu:2010fb,2011ApJ...743...28K,2011ApJ...739...52D}.
In
order to improve on these constraints significantly, direct
observations of CMB polarization are required.
Thus far the best limit from CMB polarization alone is $r<0.72$ at
95\% C.L.~\citep{Chiang:2010}, while many experiments have observed
even-parity patterns ($E$-modes)~\citep[]{leitch:2005, montroy:2006,
sievers:2007, Wu2007, bischoff:2008, Brown:2009uy,larson:2010,quiet:2011}.
Experiments currently in operation or under
construction seek to reach $r \sim 0.01$ as well as to measure the
signature of the gravitational lensing~\citep[]{2009AIPC.1185..494E,
2010SPIE.7741E..51N, 2010SPIE.7741E..40O,
CLASS.SPIE.2012,
2004SPIE.5543..320O, 2010SPIE.7741E..50S,
2010SPIE.7741E..49B, 2011AA...536A...1P,
POLAR.SPIE.2012,
2010SPIE.7741E..39A,
2008SPIE.7010E..79C, 2009AIPC.1185..511M}.

The Q/U Imaging ExperimenT (QUIET) observed the CMB from the ground
between 2008 October and 2010 December.
The observation site was the Chajnantor plateau
at an altitude of 5080\,m in the Atacama Desert in Chile. Two
different receivers were employed, corresponding to center
frequencies of 43 (Q-band) and 95\,GHz (W-band). The results of the
43-GHz measurements have been published in \citet{quiet:2011} and
included a measurement of the $E$-mode power spectrum between
$\ell=25$ and 475 and an upper limit on the tensor-to-scalar ratio of
$r<2.2$ at 95\% C.L. In this paper, we report measurements of the CMB
polarization power spectra for the 95-GHz data.
We note that this experiment played the role of a pathfinder, demonstrating
that monolithic-microwave-integrated-circuit (MMIC) arrays are capable of controlling systematic errors and
achieving the sensitivity required to reach $r \lesssim 0.01$.

QUIET was led by Bruce Winstein, who died in 2011 February soon after
observations were completed.  His intellectual and scientific guidance
was crucial to the experiment's success.

 \section{Instrument}
 In this section, we summarize the salient features of the 95-GHz
 instrument. For further details, we refer to
 separate papers \citep[]{quiet:2011,Instrumentpaper:2012}, hereafter referred to as
 \citetalias{quiet:2011} and \citetalias{Instrumentpaper:2012},
 respectively. Additional information on the QUIET instrument is
 provided in
 \citet{bischoffphd, briziusphd,cleary:77412H,Akitoproc,monsalve:77412M,newburghphd, Lauraproc}; and \citet{reevesLTD}.

The QUIET telescope consists of a 1.4-m side-fed classical
Dragonian antenna that satisfies the Mizuguchi
condition~\citepalias{Instrumentpaper:2012}.  The Cosmic Background
Imager (CBI) telescope mount was reused for the QUIET project. It
provides three-axis motion: azimuth, elevation, and rotation about the
optical axis, called ``deck'' rotation \citep{Padin:2001df}.
The 95-GHz receiver comprises 84 polarization-sensitive radiometers
and six radiometers with differential-temperature sensitivity. 
The array sensitivity is $87\,\mu{\rm K}\sqrt{\rm s}$ to the CMB
polarization.  The instantaneous angular resolution is $11\farcm7$ in
full width at half maximum (FWHM). The telescope field of view is
roughly circular with a diameter of $\sim8^{\circ}$.

The coherent QUIET radiometers directly measure the Stokes $Q$ and
$U$ parameters~\citepalias{quiet:2011,Instrumentpaper:2012}.
 The
intensity, $I$, is also recorded by the same radiometers, but with
significantly higher noise. One of the strengths of the QUIET
design is excellent immunity to both $1/f$ noise from gain fluctuations and
instrumental spurious polarization (hereafter $I$-to-$Q/U$ leakage).
The median $1/f$ knee frequency of the radiometers is
10\,mHz, significantly below the typical scan frequency of 45--100\,mHz,
resulting in a negligible $1/f$ noise contribution.
The fractional $I$-to-$Q/U$ leakages are 0.2\% for the monopole
component, 0.4\% for the dipole component, and 0.2\% for the
quadrupole component~\citepalias{Instrumentpaper:2012}.
 
The receiver and telescope mirrors are surrounded by an absorptive ground
screen, eliminating major contributions from the 300-K ground
emission.  The upper component of the ground screen was installed in
2010 January and eliminated two localized far sidelobes with
intensities $\sim -60$\,dB~\citepalias{Instrumentpaper:2012}, which
existed during the first few months of operation (from 2009 August
through 2010 January).  For the data from the early part of the
season, we reject the part where the Sun entered either of these
sidelobes.
Scan-synchronous signal due to ground emission is
projected out of the maps in the analysis~\citepalias{quiet:2011}.
Possible remaining effects are estimated as a systematic error 
(Section~\ref{sec:syst_errors}).

 \section{Observations}

With the 95-GHz receiver, we observed from 2009 August 12 until 2010
December 22 and accumulated 7426\,hours of data\footnote{  
The instrument was in the nominal CMB observing configuration only
between 2009 August 15 and 2010 December 17. Different configurations
were used between 2009 August 12 and 15 and between 2010 December 17 and
22 to calibrate and characterize the instrument.}.
Of these data, 72\%
were spent on CMB observations, 14\% on Galactic fields\footnote{The
  analysis of the Galactic observations is in progress (see
  \citet{wehus:2012} for preliminary maps), and final results will
  appear in a future publication.}, 13\% on calibration sources, and
1\% on incomplete observations due to obvious instrumental problems such as a lack of telescope
motion.  We observed 24\,hours per day, 
except for interruptions due to a variety of factors such as high wind,
heavy snow, power outages, and instrumental failures.
Our
full-season operating efficiency was 63\%. For the CMB measurements, we
selected four low-foreground sky fields, denoted CMB-1, 2, 3 and
4~\citepalias{quiet:2011}. In total, we collected 5337\,hours of CMB data
with the 95-GHz receiver (Table~\ref{tab:selection}).
\begin{deluxetable}{lcccc}
\tablecaption{Data-selection Summary\label{tab:selection}}
\tablecolumns{7}
\tablehead{ & Observed Time & \multicolumn{3}{c}{Data Percentage} \\
  Field & (Hours) & ML & PCL & Both}
\tablecomments{ Fraction of data selected for each field by
  each pipeline.  The last column shows the fraction simultaneously selected
 by both pipelines. 
}
\startdata
CMB-1 & 1\,855 & 69.7 & 64.4 & 57.7 \\
CMB-2 & 1\,444 & 73.1 & 67.1 & 61.2 \\
CMB-3 & 1\,389 & 64.4 & 58.8 & 52.6 \\
CMB-4 & \phantom{1}\phantom{,}650 & 72.1 & 65.4 & 60.4 \\
Total &5\,337  & 69.5 & 63.5 & 57.6
\enddata
\end{deluxetable}

Each observation consists of a series of
constant-elevation scans, hereafter collectively called a CES.  The scans are in the azimuth
direction
with a half amplitude of $7.5^{\circ}$ on the sky.
Diurnal motion of the sky causes the field to drift through the field
of view. After the target has drifted $15^{\circ}$ on the sky, we
adjust the azimuth and elevation to retrack the field and begin a new
CES.
Each
individual CES thus scans over an area of
$\sim15^{\circ}\times15^{\circ}$.  Due to the field of view of
$\sim8^\circ$ and the fact that the sky does not always drift
orthogonal to the scan direction, a larger area is observed in practice.
The deck angle is changed by
$45^{\circ}$ each week, providing a large degree of immunity to
spurious $B$-modes induced by $I$-to-$Q/U$ leakage.

 \section{Calibration}

The instrument calibration procedure for the 95-GHz observations is
similar to that used for the 43-GHz
data~\citepalias{quiet:2011,Instrumentpaper:2012}.
 The instantaneous beam point-spread
function is derived from observations of Taurus A (hereafter
Tau~A). The resulting beam function has a width of $11\farcm7$
FWHM with a small non-Gaussian correction~\citepalias{Instrumentpaper:2012}.
The telescope pointing model is calibrated with a set of 
astronomical objects: Tau~A, Jupiter, RCW~38,
the Moon, and the Galactic center. The residual random scatter after
applying all pointing corrections is $5\farcm1$ FWHM.
To correct for this, we convolve the beam
window function with the residual--pointing-scatter term, and obtain an
effective point-spread function of $12\farcm8$ FWHM.
The detector angles (i.e., the orientations of the polarization
responses) are calibrated to $0\fdg5$ precision with the
combination of Tau~A observations for absolute-angle determination and
a sparse--wire-grid
calibrator~\citepalias[\citealt{tajimaltd};][]{Instrumentpaper:2012} for relative angle
determination. The considerable improvement in the detector angle
precision compared to the previous 43-GHz analysis~\citepalias{quiet:2011}
is due to a more accurate catalog value of Tau A~\cite[]{IRAM/TauA} as well as
an improved wire grid calibration.
Large and small sky dips (elevation
nods of $\pm 20^{\circ}$ and $\pm 3^{\circ}$ amplitudes,
respectively) modulate loading from atmospheric emission and
allow us to measure the fractional
$I$-to-$Q/U$ monopole leakage with 0.3\% precision per calibration,
while Jupiter measurements are used to measure the higher-order
leakage terms (i.e., dipole and quadrupole) and to confirm the sky-dip monopole-leakage results.
The detector responsivities are calibrated using Tau A and sky-dip
data as well as the measurement using the sparse wire grid. The typical
responsivity is found to be 3.1\,$\mathrm{mV\,K^{-1}}$ in antenna
temperature units.

\section{Data Analysis}

The analysis procedure used for the 95-GHz data reduction
follows closely the 43-GHz analysis~\citepalias{quiet:2011}, and we refer
the reader to this publication as well as recent Ph.~D. theses
\citep{buderphd,chinonephd,dumoulinphd,monsalvephd,naessphd} for full details. We have
implemented two independent analysis pipelines, one based on a
maximum-likelihood (ML) technique and the other on a pseudo-$C_\ell$
(PCL) cross-correlation technique. The most important improvements
since the previous publication are, for the ML pipeline, an adaptive
filter procedure in which the filter parameters depend on the data
quality of the specific data segment, as well as a pseudo-$C_{\ell}$
null-test estimator, allowing for many more null tests; and, for the
PCL pipeline, a different and more robust data division for the
cross-correlation\footnote{We cross-correlate among 40 subsets of data.
Each subset corresponds to a specific boresight azimuth and deck range.
There are five azimuth and eight deck ranges.}, taking further advantage
of the scanning strategy.

\begin{figure*}[thb]
\centering
\includegraphics[width=0.9\textwidth]{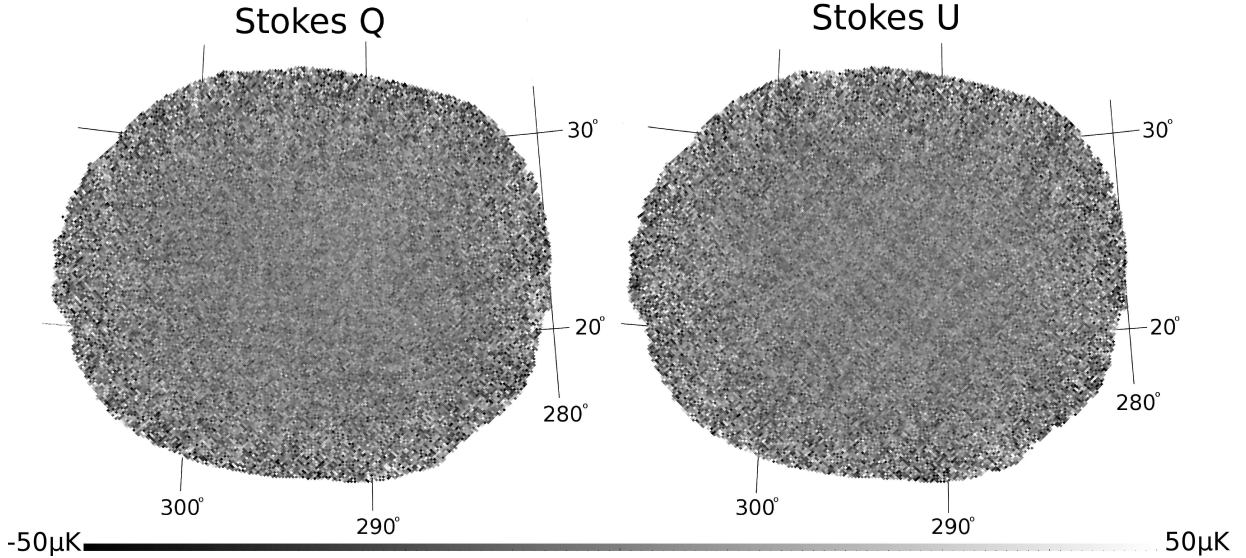}
\caption{
QUIET CMB polarization maps of the CMB-1 field in
 Galactic coordinates at
95\,GHz. The left (right) panel shows Stokes $Q$ ($U$),
 where the polarization angle is defined with respect to the Galactic North
 Pole.
 Note the
coherent vertical/horizontal patterns in the $Q$ map, and the
diagonal patterns in the $U$ map; these are the expected signature of a
pure $E$-mode signal. No filtering has been applied to this map beyond
subtracting the very largest angular scales ($\ell < 25$), to which
 QUIET is not  sensitive. 
}
 \label{fig:map}
\end{figure*}

The process to extract cosmological results from raw
time-ordered data (TOD), containing 
measurements of the Stokes $Q$ and $U$ parameters as well as the
telescope pointing information, can be
summarized in three steps: TOD pre-processing, map making, and power-spectrum and parameter estimation.
The TOD pre-processing involves
estimating and applying calibration factors, characterizing the
detector noise, and applying high-pass, low-pass and azimuth filters
to minimize the effects of atmospheric fluctuations, far sidelobes,
excess high-frequency instrumental noise, and ground pickup.
Then, sky maps are generated by projecting the $Q$ and $U$ intensities
into Galactic coordinates, taking into account the telescope pointing
information, using standard map-making equations
\citepalias{quiet:2011}.  Figure~\ref{fig:map} shows the
maximum-likelihood Stokes $Q$ and $U$ maps of the CMB-1
field generated with the ML pipeline. 
Power-spectrum estimation is performed with one of two techniques,
depending on the pipeline. The ML pipeline implements a standard
Newton--Raphson maximum-likelihood solver \citep{Bond1998}, while the
PCL pipeline implements the MASTER pseudo-$C_{\ell}$ algorithm
\citep{Hivon2002,Hansen:2002iha}. Prior to power-spectrum
estimation, both pipelines mask Centaurus A, and the PCL pipeline also
masks Pictor A.

In the following we describe the data selection, analysis validation
and systematic-error assessment.
In optimizing the analysis configuration, it is important that the
optimization process itself does not introduce experimenter biases,
for instance by removing purely statistical fluctuations in the data
selection.
QUIET is the first CMB experiment to have adopted a strict blind-analysis policy \cite[]{2005ARNPS..55..141K}, in which all data-selection criteria, filters, and calibrations are adjusted and finalized,
and the systematic errors are assessed prior to looking at any
cosmological power spectrum. This process was described in 
detail in \citetalias{quiet:2011}, and we have adopted the same policy for
the 95-GHz analysis.

\subsection{Data Selection}

Each QUIET radiometer provides four output channels (``detector
diodes''), resulting in a total of 336 output channels from 84
polarization-sensitive radiometers. We use 308 good channels for
analysis~\citepalias{Instrumentpaper:2012}. Starting from the resulting
data for all CESes, we define two different classes of data-selection
criteria. 
In the first class, we impose criteria that select or reject
an entire CES. These
include the criteria based on atmospheric conditions, instrument
malfunctions, or unusual conditions for the temperature
regulation in the focal plane.
In the second class, we apply selection criteria to individual
detector diodes in each CES (CES-diodes). For instance, a CES-diode is rejected
if: 1) the measured noise
properties show poor agreement with the noise model; 2) the $1/f$ knee frequency
is anomalously high; 3) the white-noise level is
non-stationary; 4) there are glitches in the time domain
or strong spikes in the Fourier domain; or 5) there is
evidence of a large scan-synchronous signal.

Table~\ref{tab:selection} lists the fractions of data that satisfy the
criteria and are used for map making and power-spectrum
estimation.

 \subsection{Analysis Validation}
Having defined our data-selection criteria and filters, we need to
validate the accepted data set and analysis parameters\footnote{
 Note that the data-selection criteria are improved through an iterative
 process of applying the analysis-validation metrics.
  In the current analysis, $\sim50$
  different configurations were considered before reaching the final
  configuration.}.  Our most valuable tool for this is a so-called
null-test suite~\citepalias{quiet:2011}.  In each null test, the full
data are split into two subsets. From these,
we make individual sky
maps, $m_1$ and $m_2$, as well as the corresponding difference map,
$m_{\textrm{diff}} \equiv ( m_1 - m_2 ) / 2$. 
By design, the true sky signal cancels in this map, and the result
should be consistent with noise.
We therefore compute the $EE$ and $BB$ power spectra of this map, and
check for consistency with the zero-signal hypothesis by comparing to
simulations. In the current analysis, the null suite consists of 32
and 23 tests for the PCL and ML pipelines, respectively, with each test
targeting a possible source of signal contamination or miscalibration.
These are selected to be highly independent;
a statistical correlation between null power spectra of two different null-test divisions is typically 0.05.

For each power-spectrum bin $b$, we calculate the statistic $\chi_{\rm
  null}(b) \equiv C^{\rm null}_{b}/\sigma_{b}$, where $C^{\rm null}_{b}$
is the observed difference power spectrum and $\sigma_{b}$ is a
Monte Carlo (MC) based estimate of the corresponding standard deviation.
We evaluate both $\chi_{\rm null}$ and its square for all $b$; $\chi_{\rm null}$
is sensitive to systematic biases in the null spectra, while $\chi_{\rm
  null}^2$ is more responsive to outliers.

\begin{figure}[tb]
\includegraphics[width=0.48\textwidth,clip=]{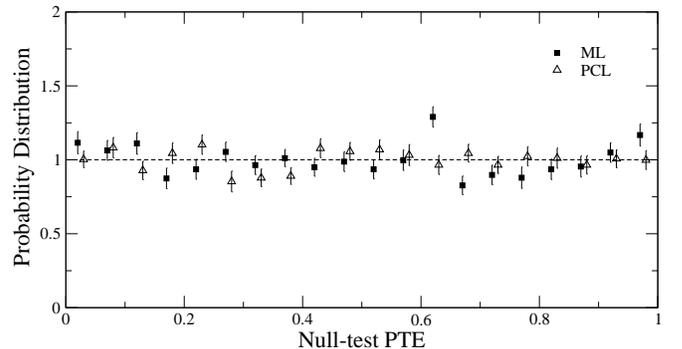}
\caption{Null-test--PTE distributions for $\chi_{\textrm{null}}^2$ for
  both the ML and PCL pipelines. Each 
  is consistent with the uniform expectation.   }
 \label{fig:null}
\end{figure}

Prior to the analysis, we defined three critical tests that had to
be passed before continuing to cosmological analysis, based on 1) the
mean value of $\chi_{\rm null}$, 2) the sum of $\chi_{\rm null}^2$,
and 3) the maximum of $\chi_{\rm null}^2$, all computed including the
entire suite of $EE$ and $BB$ null power spectra. A given analysis
configuration passes when these
statistics are consistent with the null hypothesis.
Table~\ref{tab:validation} lists the probabilities to exceed (PTE)
for the final configuration,
and Figure~\ref{fig:null} shows the PTE distribution of $\chi_{\rm
  null}^2$.
The PTEs are defined such that a large deviation from zero
 results in a low PTE.
 This corresponds to two-sided PTEs for the mean of $\chi_{\rm null}$
 and one-sided PTEs for the total $\chi_{\rm null}^2$
 and the $\chi_{\rm null}^2$ outlier.
The mean of the $\chi_{\rm null}$ distributions over all
fields is $-0.018 \pm 0.015$ for the PCL pipeline and $0.003\pm0.017$
for the ML pipeline.
We do not detect any bias with our final analysis
configuration.
\begin{deluxetable}{lcccccc}
\tablecaption{Validation-test Summary\label{tab:validation}}
\tablecolumns{7}
\tablehead{ & \multicolumn{2}{c}{Mean of $\chi_{\rm null}$}
 & \multicolumn{2}{c}{Total $\chi_{\rm null}^2$}
 & \multicolumn{2}{c}{$\chi_{\rm null}^2$ Outlier} \\
Field & ML & PCL & ML & PCL & ML & PCL}
\tablecomments{ Results of the three predefined validation
 tests  using
 the mean of $\chi_{\rm null}$, the sum of $\chi_{\rm null}^2$,
 and the worst outlier of $\chi_{\rm null}^2$.
 All values are PTEs defined such that a large deviation from 
 zero results in a small PTE.
 }
\startdata
CMB-1 & 0.40 & 0.26 & 0.31 & 0.78 & 0.63 & 0.14 \\
CMB-2 & 0.54 & 0.46 & 0.08 & 0.06 & 0.02 & 0.40 \\
CMB-3 & 0.42 & 0.74 & 0.31 & 0.50 & 0.72 & 0.45 \\
CMB-4 & 0.06 & 0.08 & 0.43 & 0.76 & 0.80 & 0.18
\enddata
\end{deluxetable}

We also generate 1000 random null divisions and compare the
widths of the resulting $\chi_{\mathrm{null}}$ distributions between
data and MCs using the PCL pipeline. We find these to be
consistent, and we verify our estimate of the statistical
uncertainty in each multipole bin with a precision of 3.1\%.
Finally, we evaluate the differences of non-null spectra among the
fields, before looking at individual non-null spectra.  These
differences are consistent with the hypothesis of statistical isotropy 
(i.e., each field has the
same underlying power spectrum), with a PTE of 0.15.

 \subsection{Systematic Errors}
\label{sec:syst_errors}
\begin{figure*}[thb]
\includegraphics[width=\textwidth,clip=]{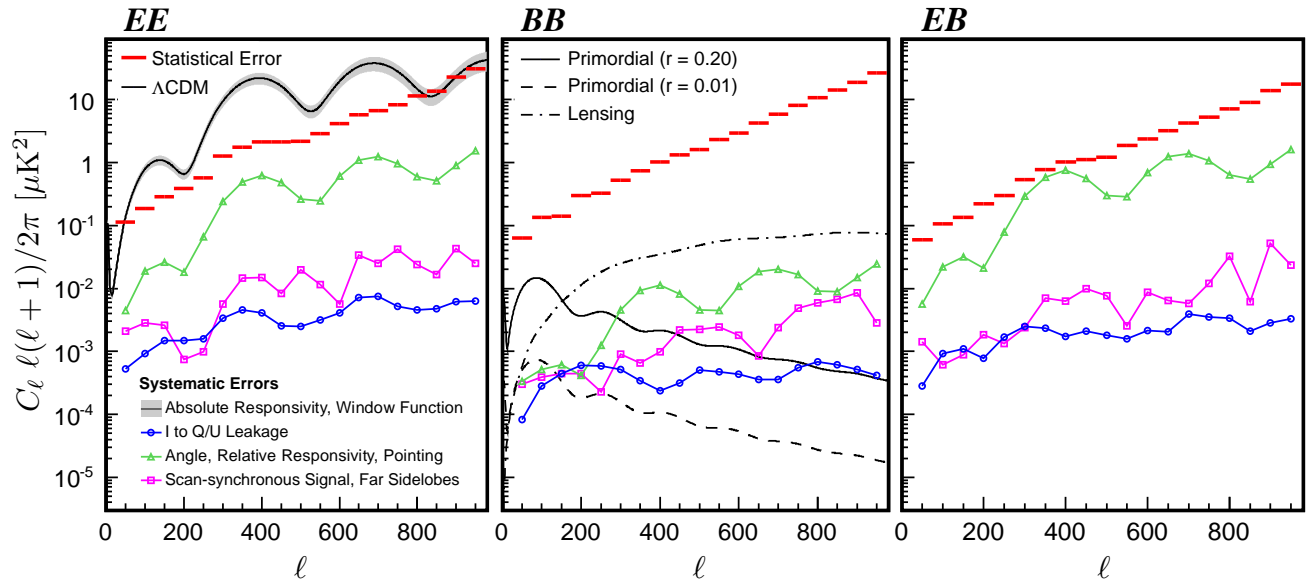}
\caption{Summary of systematic error assessment for $EE$ (left), $BB$
  (middle), and $EB$ (right). The red bars indicate the
 statistical uncertainties in each bin.
 Blue, green, and purple points correspond to three categories of
 systematic errors: $I$-to-$Q/U$ leakage; polarization angles (absolute
 and relative), relative responsivities and pointing error; and the
 residual scan-synchronous signals and far sidelobes.  The gray band along
 the $\Lambda$CDM curve in $EE$ corresponds to the uncertainties of
 multiplicative factors: absolute responsivity and the window
 function.
 For $BB$, all systematic errors are
  below the level of $r \sim0.01$ at $\ell \sim100$.
 For $EE$ the dominant systematic
  error is uncertainty in the absolute responsivity, which is a purely
multiplicative effect. For $EB$, the dominant systematic is caused by
uncertainties in the polarization detector angle. }
 \label{fig:syst_errors}
\end{figure*}

We study the contributions from instrumental systematic errors using the 
methodology of \citetalias{quiet:2011}. The main effects
considered are 1) uncertainties
in absolute responsivity and the window function, 2) $I$-to-$Q/U$
leakage, 3) uncertainties in polarization angles, relative
responsivities, and pointing, and 4) residual contamination from
scan-synchronous signals and far sidelobes. In each case, we set up an
empirical model of the systematic effect and propagate this through the
PCL pipeline. The results from these calculations are summarized
in Figure \ref{fig:syst_errors}.

The most important conclusion is that the systematic errors in the
$BB$ spectrum are very small. For the multipole range relevant for
estimation of the tensor-to-scalar ratio, $\ell \sim 100$, each
effect is smaller than or comparable to the signal corresponding to
$r\sim0.01$, the lowest level ever reported in the literature. It is
also noteworthy that this limit improves on that reported for the
43-GHz data ($r < 0.1$; QUIET 2011) by an order of magnitude. This is
due to improved rejection of $I$-to-$Q/U$ leakage, better detector-angle 
calibration, and lower levels of sidelobe contamination
resulting from the installation of the upper parts of the ground
screen.

For the $EE$ power spectrum, the systematic error budget is
dominated by uncertainties in the multiplicative responsivity
calibration. The total uncertainty is 8\%, almost equally contributed
from three dominant sources: the uncertainty of the
polarization flux of Tau A \cite[5\%;][]{Weiland2010}, the
uncertainty in the beam solid angle (5\%), and the uncertainty
associated with modeling the time variation and relative responsivity
among the detector channels (4\%). This translates into an uncertainty
of 17\% in the power spectrum. For comparison, the statistical
uncertainty in the $EE$ spectrum is about 8\% of the central value
at its minimum around $\ell\sim400$. It is important to note
that the responsivity
effect is purely
multiplicative and therefore cannot create spurious $B$-mode
signal.
The uncertainty of the window function
is another multiplicative factor highly correlated among 
different $\ell$ bins, with the magnitude dependent on $\ell$.
The uncertainty comes from both the beam window function and the
smearing factor due to the pointing error, and is listed in
Table~\ref{tab:spectra2}.  These errors are smaller than the
$EE$ statistical uncertainties.

The dominant systematic uncertainty for $EB$ is due to calibration
errors in the detector polarization angles.
 To first
approximation, an error in the absolute--polarization-angle calibration
of $\delta \psi$ induces a spurious $EB$ spectrum proportional to
$\sim C_\ell^{EE} \sin 2\delta \psi$, and a $BB$ spectrum proportional
to $\sim C_\ell^{EE} \sin^2 2\delta \psi$. 
Uncertainties in the relative polarization angles among detectors
contribute to the systematic errors in $EB$ and $BB$ spectra in a
similar manner.  The calculations summarized in Figure
\ref{fig:syst_errors} capture both these effects through
simulations based on the $\Lambda$CDM prediction for
$C_\ell^{EE}$. As seen in this figure, these polarization-angle
uncertainties lead to systematic errors almost as large as the
statistical errors for $EB$ around $\ell\sim400$, while for $BB$ they
are small everywhere and comparable to other sources of systematic
errors.
Table~\ref{tab:spectra2} lists the total systematic error for the $EB$ power
spectrum.

 \section{Power spectra and cosmological parameters}

\begin{figure}[thb]
\includegraphics[width=0.45\textwidth]{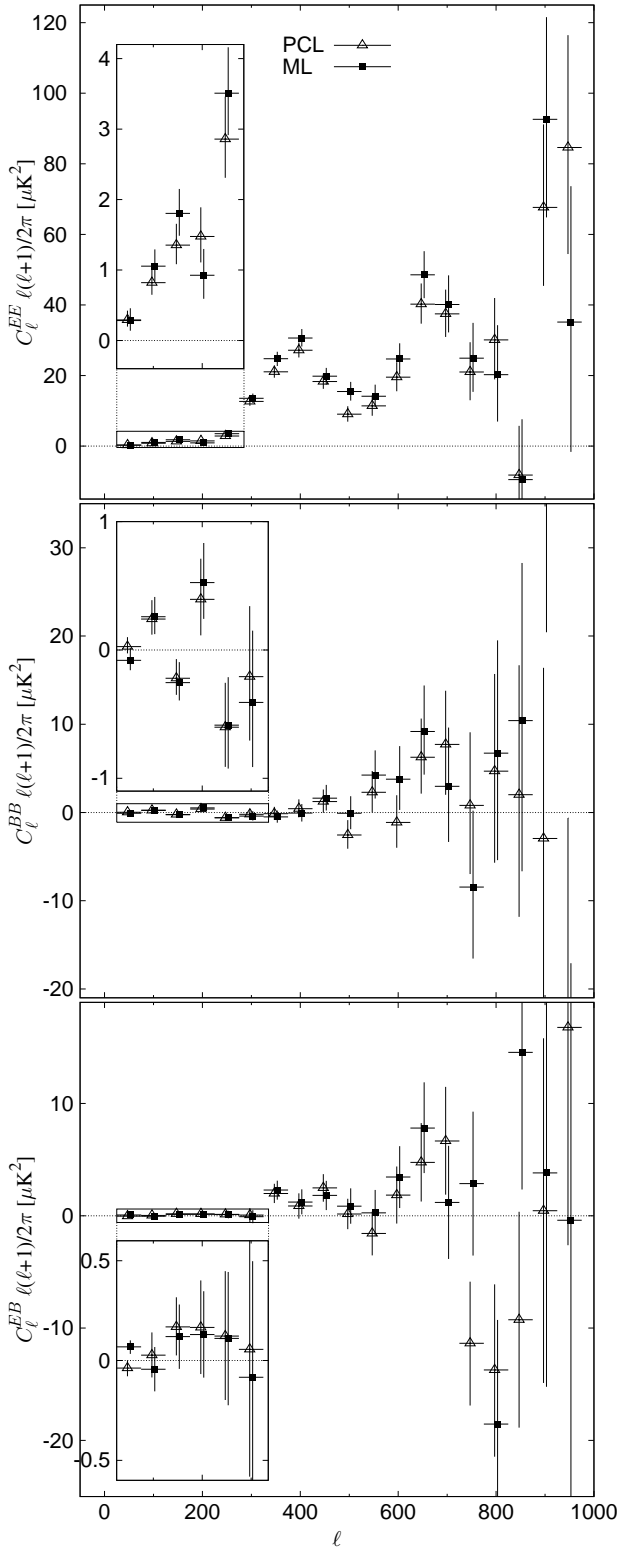}
\caption{ The QUIET 95-GHz power spectra, co-added over all four CMB
  fields. The panels show the $EE$ (top), $BB$ (middle) and $EB$
  (bottom) spectra, and the insets show the low-$\ell$ region in
  detail.
  The central $\ell$ values from the two pipelines are slightly offset for display purposes.
 Note that the error bars indicate statistical errors only;
  see Section \ref{sec:syst_errors} for a discussion of systematic
  errors.
 Typical correlations among neighboring bins are $\sim -0.1$.
 The full set of three spectra are consistent 
 with the $EE$ spectrum predicted by the $\Lambda$CDM model
 and $C_\ell^{BB} = C_\ell^{EB} = 0$.
 }
 \label{fig:cl_wband}
\end{figure}

The measurements of the $EE$, $EB$ and $BB$ power spectra are
tabulated in Table \ref{tab:spectra2}, and plotted in Figure
\ref{fig:cl_wband}.
The $EE$ spectrum is strongly
signal-dominated up to $\ell \sim 800$, and three acoustic peaks
are clearly traced.
Both the $BB$ and $EB$ spectra are consistent with zero within the
estimated statistical and systematic uncertainties.
The dominant $EE$ power is also visible in the maps shown in
Figure \ref{fig:map}. 
Note that these maps have not been filtered,
except by subtracting the very largest scales ($\ell \lesssim 25$),
to which QUIET is not sensitive.  One can see a distinct
vertical-horizontal coherent pattern on small angular
scales in the Stokes $Q$ map, and a similar diagonal pattern in the $U$
map.
This is the expected signature
of an $E$-mode signal.

The results from the two pipelines are consistent with each other. The
most noticeable difference is a single overall multiplicative factor,
which is only relevant in evaluating the consistency of the $EE$ power
spectra. This factor comes from different responsivity modeling and is
consistent with the systematic error budget discussed in Section
\ref{sec:syst_errors}.

\begin{figure}[tp]
\includegraphics[width=0.45\textwidth]{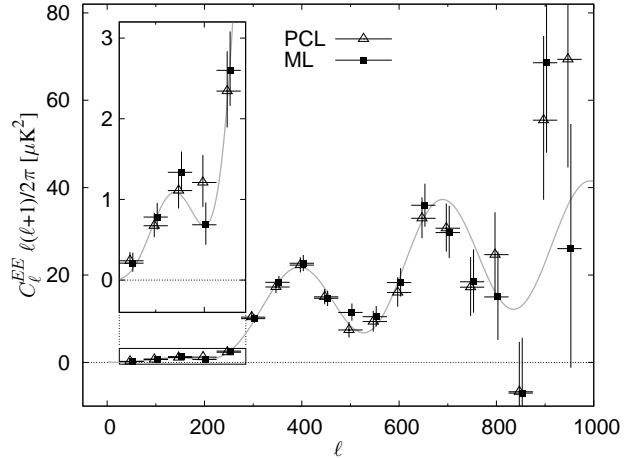}
\caption{Comparison of the QUIET $EE$ spectrum with current
  best-fit $\Lambda$CDM model after scaling the absolute
  responsivity to $q=1$.
  The central $\ell$ values from the two pipelines are slightly offset for display purposes.
 Typical correlations among neighboring bins are $\sim -0.1$.
 The results from the two pipelines are consistent, and
  the shape of the QUIET $EE$ spectrum is in excellent agreement with
  the $\Lambda$CDM model. 
 }
 \label{fig:cl_wband_scaled}
\end{figure}

When assessing the consistency of the $EE$ power spectrum with the
$\Lambda$CDM prediction, it is convenient to factor the spectrum
measurement into an overall amplitude and the spectral shape of the
acoustic peaks. We fit a free amplitude, $q$, relative to the $EE$
spectrum predicted by the best-fit seven-year \textit{WMAP} $\Lambda$CDM parameters
\citep{Komatsu:2010fb} to the spectrum from each pipeline, and find
$q=1.22 \pm 0.04 \textrm{(stat)} {}^{+0.22}_{-0.17} \textrm{(syst)}$
and $q=1.35 \pm 0.05 \textrm{(stat)}
{}^{+0.26}_{-0.22}\textrm{(syst)}$ for the PCL and ML pipelines,
respectively.
These values are consistent with the 
 $\Lambda$CDM prediction of $q=1$, and correspond to PTEs of 0.20 and
 0.06, respectively.
Figure~\ref{fig:cl_wband_scaled} provides a spectral shape comparison.
Here we see that the measured $EE$ spectrum rescaled to $q=1$
accurately traces the first three acoustic peaks
predicted by the $\Lambda$CDM model.
We assess the overall consistency with the $\Lambda$CDM hypothesis by
calculating a total $\chi^2$ relative to $\Lambda$CDM (and relative to
$C_\ell^{BB} = C_\ell^{EB} = 0$), taking into account the systematic
uncertainties due to the responsivity calibration in $EE$ and the
systematic error in $EB$ primarily due to detector polarization
angles.
The former is incorporated by introducing a nuisance parameter for the
absolute responsivity constrained by a Gaussian distribution with a standard deviation
equal to the assigned systematic error.  The latter is incorporated by
modeling
the $EB$ systematic error as $s C_\ell^{EB,\mathrm{syst}}$, where the
scale factor
$s$ is constrained by a Gaussian with $\sigma=1$ and
$C_\ell^{EB,\mathrm{syst}}$ is
the systematic error estimated in Section~\ref{sec:syst_errors}; this
means we assume the
systematic errors are completely correlated among different $\ell$ bins. 
The systematic errors in $BB$ are negligibly small.
Including
the systematic error contributions,
we find $\chi^2$ of 67.3 and 67.9 for the PCL and
ML pipelines, respectively. With 57 degrees of freedom, these values
correspond to PTEs of 0.16 and 0.15, respectively; the derived
spectra are consistent with $\Lambda$CDM.

Since we find no significant excess in the $BB$ power spectrum, we
place an upper limit on possible $BB$ power in each bin.  The ML pipeline
calculates the upper limit by the 95\% integral of the positive part
of the likelihood, while the PCL pipeline adopts a frequentist-based
hypothesis-testing method. Specifically, the upper limit $\mu$ is
defined by
$0.05=p(q_\mu>q_\mu^{\mathrm{obs}}|\mu)/p(q_\mu>q_\mu^{\mathrm{obs}}|0)$,
where $p(\cdots|\mu)$ and $p(\cdots|0)$ represent $p$-values of the null
hypothesis with power $\mu$ and an alternative hypothesis with zero
power, respectively. The parameters $\mu$, $q_\mu$ and $q_\mu^{\mathrm{obs}}$
correspond to the bandpower $C_b$, the test statistic for upper limit
defined in \cite{2011EPJC...71.1554C}, and the test statistic $q_\mu$
calculated for the observed bandpower $\hat{C}_b$, respectively.
Table~\ref{tab:spectra2} lists the derived upper limits.

 We constrain the tensor-to-scalar ratio $r$ using standard
 likelihood methods and including only the $BB$ spectrum at low
 multipoles ($26\leq \ell \leq 175$). For simplicity, we consider only
 the amplitude of a $BB$ template computed with the standard
 $\Lambda$CDM concordance parameters, and fix the tensor spectral
 index to $n_t = 0$~\cite[\citetalias{quiet:2011};][]{Chiang:2010}.
 In constraining $r$,
 the uncertainty of the responsivity calibration is eliminated
 by simultaneously fitting $EE$ and $BB$ power spectra using the
 $\Lambda$CDM templates.  We define the fit function as
 $C_{\ell}^{EE}(q) = q\, C_{\ell}^{EE,\textrm{fid}}$ and
 $C_{\ell}^{BB}(q,r) = r\,q\, C_{\ell}^{BB,\textrm{fid}}$.  Here
 $C_{\ell}^{EE,\textrm{fid}}$ and $C_{\ell}^{BB,\textrm{fid}}$ denote
 the fiducial $\Lambda$CDM $EE$--power-spectrum template and the 
 $BB$--power-spectrum template with $r=1$ and $n_t=0$, respectively.  Note
 that this does not imply that we use $EE$ power to constrain the
 tensor modes, as the $\Lambda$CDM $EE$ template only contains the scalar
 contribution. This method exploits the fact that $r$ is by definition a
 ratio and does not depend on the common overall scaling factor. From
 the simultaneous fit, the ML pipeline finds $r = 1.1^{+0.9}_{-0.8}$,
 with a 95\% C.L. upper limit of $r<2.8$, and the PCL pipeline finds
 $r = 1.2^{+0.9}_{-0.8}$, corresponding to an upper limit of
 $r<2.7$. The systematic uncertainty is negligible, at the level of
 $r=0.01$.

\section{Foregrounds}
We assess the level of diffuse foregrounds, in particular synchrotron
radiation and dust emission, as additional sources of systematic
errors. Contamination from residual point sources is negligible.  An
estimate using the point-source component of the Planck Sky Model
~\cite[PSM;][PSM v1.7.4]{PlanckSkyModel} yields a limit of
$C_\ell < 1.4\times 10^{-6}\,\mu{\rm K}^2$ over the entire $\ell$ range
 without masking any sources\footnote{Note that
  this limit is given in units of $C_\ell$, not
  $C_{\ell}\,\ell(\ell+1)/2\pi$.}. An estimate based on a source-population 
model~\cite[]{2012arXiv1204.0427T} relative to our nominal
point-source mask results in an even lower level, $C_\ell \sim 5\times
10^{-7}\,\mu{\rm K}^2$.  Both are well below our statistical
uncertainty.

Considering synchrotron radiation, we note that the 43-GHz QUIET
observations have already resulted in strong constraints on any
synchrotron component in each of the QUIET CMB fields
\citepalias{quiet:2011}.  Except for the single case of the $EE$ spectrum
at $\ell \le 75$ measured in CMB-1, no evidence of any contamination
was found. These results allow us to constrain any contribution from
synchrotron emission at 95\,GHz by extrapolation.  
Adopting a spectral index of $\beta_{\textrm{s}} = -2.7$
\citep{dunkley:2009}, we estimate the $EE$ ($BB$) excess power to be
$0.011\pm0.003\,\mu{\rm K}^2$ ($0.001\pm0.002\,\mu{\rm K}^2$) for the
first bin of the CMB-1 spectrum, which is negligible compared to
statistical errors.

In order to constrain contamination from dust emission, we adopt the
thermal-dust component of the PSM as a
template; the PSM predicts that other sources of contamination are
subdominant at 95\,GHz in the QUIET fields.
We estimate the dust power contribution in our fields by
evaluating both the PSM power spectrum and the PSM-QUIET
cross-spectrum using the PCL pipeline.
The possible contamination is only relevant in the first bin ($25 \leq
\ell \leq 75$) of the field CMB-1.  In this bin, the PSM power
amplitude is $0.087\,\mu\textrm{K}^2$ ($0.070\,\mu\textrm{K}^2$) for the
$EE$ ($BB$) spectrum, while the corresponding cross power is
$0.060\pm0.035\,\mu\textrm{K}^2$
($0.016\pm0.027\,\mu\textrm{K}^2$). Taking into account the relative
weights of the individual fields, we therefore estimate that the dust-emission
 contribution to the first $EE$ bin in the final co-added
spectrum (Table \ref{tab:spectra2}) is $<0.04\,\mu\textrm{K}^2$, more than
a factor two smaller than the statistical uncertainty. All other spectra
and multipole ranges have negligible contributions.
Fitting the PSM model as a template to CMB-1 in the map domain using the
ML pipeline, we find a best-fit amplitude of $A = 0.62\pm0.21$. This
corresponds to a $3\,\sigma$ correlation with the thermal-dust PSM
component, which at the same time agrees with the PSM prediction
($A=1$) at $1.8\,\sigma$. Consistent results are obtained by taking the
ratio of the cross-power to the PSM power including the full multipole
range, with an amplitude of $A = 0.66 \pm 0.25$. The three other
fields all have best-fit amplitudes consistent with zero. 
We note as a caveat that the uncertainty in the PSM itself is not
taken into account in this analysis, and the results depend
critically on this model as the detected foreground levels are well
below the statistical errors of the measured power spectra themselves.

 \section{Conclusions}
\begin{figure}[!t]
\includegraphics[width=0.47\textwidth,clip=]{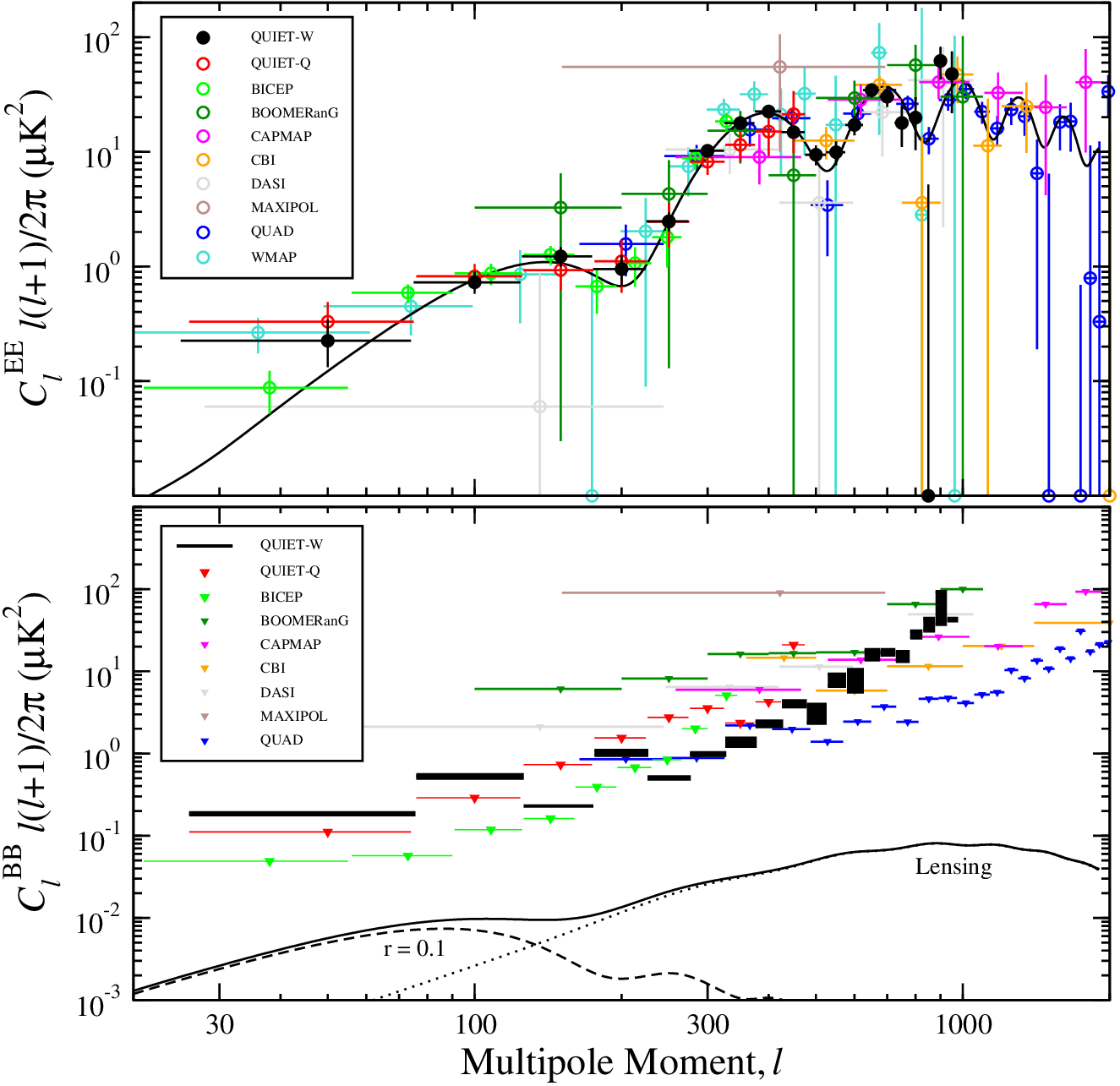}
\caption{ Summary of published CMB polarization $EE$ power spectrum
  (top) and 95\% C.L. upper limits on $BB$ power (bottom) measured by
  different experiments \citepalias[\citealt{leitch:2005,
  montroy:2006, sievers:2007, Wu2007, bischoff:2008,
  Brown:2009uy, Chiang:2010, larson:2010};][]{quiet:2011}
 as well as the result reported in this paper (QUIET-W).
 The QUIET-W points, spanning the first three acoustic peaks in the $EE$
 power spectrum, bridge the large ($\ell \lesssim 200$) and small
 ($\ell \gtrsim 400$) angular-scale measurements made by previous
 experiments.
 For visualization
 purposes, the mean of two
  pipeline spectra (scaled to $q=1$) is shown for QUIET-W for
  $EE$.  
 For $BB$, the results from the two individual pipelines are
  indicated by the vertical extent of the QUIET-W points. The solid
  line in the upper panel shows the $\Lambda$CDM $EE$ spectrum; the
  dashed and dotted lines in the bottom panel show the $BB$ spectrum
  from gravitational waves (for $r=0.1$) and lensing, respectively. }
 \label{fig:experiments}
\end{figure}

We have presented the CMB polarization power spectra from the 95-GHz QUIET
observations.  The $EE$ spectrum has been measured between $\ell=25$
and $975$, and the first three acoustic peaks were seen with high
signal-to-noise ratio, consistent with $\Lambda$CDM predictions. The
$BB$ spectrum was found to be consistent with zero, with a 95\%
C.L. upper limit on the tensor-to-scalar ratio of $r<2.7$ (PCL) or 2.8 (ML),
depending on pipeline. In Figure \ref{fig:experiments}, we provide an
up-to-date overview of the current state of the CMB polarization
field, comparing the results from various experiments\footnote{For the
  $EE$ spectrum of QUIET, we show the mean of the spectrum from the two
  pipelines (after scaling to $q=1$) as a succinct visualization. For
  $BB$, the results from the two individual pipelines are indicated by
  the vertical extent of the QUIET-W points.}.
In one of the fields, we found a correlation with the dust component of
 the Planck Sky Model.  The excess power due to this
 component was still small compared to the statistical errors of the power spectra.
  Finally, we have demonstrated
the lowest level of instrumental systematic errors to date.
We conclude by noting that part of the role of this experiment was to
serve as a pathfinder to demonstrate that MMIC arrays were capable of
reaching $r \lesssim 0.01$; this has been successfully achieved.

\begin{acknowledgements}
Support for the QUIET instrument and operation comes through the NSF
cooperative agreement AST-0506648. Support was also provided by NSF
awards PHY-0355328, AST-0448909, PHY-0551142, PHY-0855887, and
 AST-1010016; KAKENHI
20244041, 20740158, and 21111002; PRODEX C90284; a KIPAC Enterprise
grant; and by the Strategic Alliance for the Implementation of New
Technologies (SAINT).

Some work was performed on the Joint Fermilab-KICP Supercomputing
Cluster, supported by grants from Fermilab, the Kavli Institute for
Cosmological Physics, and the University of Chicago.  Some work was
performed on the Titan Cluster, owned and maintained by the University
of Oslo and NOTUR (the Norwegian High Performance Computing
Consortium), and on the Central Computing System, owned and operated
by the Computing Research Center at KEK. 
This research used resources of the National Energy Research
Scientific Computing Center, which is supported by the Office of
Science of the U.S. Department of Energy under Contract
No. DE-AC02-05CH11231. 
Portions of this work were performed at the Jet Propulsion Laboratory
(JPL) and California Institute of Technology, operating under a
contract with the National Aeronautics and Space Administration. The
Q-band modules were developed using funding from the JPL R\&TD
program.  We acknowledge the Northrop Grumman Corporation for
collaboration in the development and fabrication of HEMT-based
cryogenic temperature-compatible MMICs. We acknowledge the use of the
Planck Sky Model, developed by the Component Separation Working Group
(WG2) of the Planck Collaboration. Some of the results in this paper have been
derived using the HEALPix \citep{Gorski:2004by} software and analysis
package.

C.D. acknowledges an STFC Advanced Fellowship and an ERC IRG grant under FP7.
R.B. acknowledges support from CONICYT project Basal PFB-06 and
ALMA-Conicyt 31070015.
A.D.M. acknowledges a Sloan foundation fellowship. H.K.E. acknowledges an
ERC Starting Grant under FP7.

PWV measurements were provided by the Atacama Pathfinder Experiment
(APEX). We thank CONICYT for granting permission to operate within the
Chajnantor Scientific Preserve in Chile, and ALMA for providing site
infrastructure support.
Field operations were based at the Don Esteban facility run by Astro-Norte.
We are particularly indebted to the engineers
and technician who maintained and operated the telescope: Jos\'e Cort\'es,
Cristobal Jara, Freddy Mu\~noz, and Carlos Verdugo.

In addition, we would like to acknowledge the following people for
their assistance in the instrument design, construction,
commissioning, operation, and in data analysis: Augusto Gutierrez
Aitken, Colin Baines, Phil Bannister, Hannah Barker, Matthew R. Becker, Alex
Blein, 
Mircea Bogdan, 
Anushya Chandra, Sea Moon Cho, Joelle Cooperrider,
Mike Crofts, Emma Curry, Maire Daly, Richard Davis, Fritz Dejongh, Joy Didier, Greg
Dooley, Hans Eide, Pedro Ferreira, 
Jonathon Goh, 
Will Grainger, 
Peter Hamlington,
Takeo Higuchi, Seth Hillbrand, Christian Holler,
Ben Hooberman, Kathryn D. Huff, 
William Imbriale, 
Oliver King, Eiichiro Komatsu,
Jostein Kristiansen,
Richard Lai, 
David Leibovitch, Erik Leitch, Kelly Lepo, Siqi Li, Martha Malin, Mark McCulloch, Steve Meyer, Oliver Montes, David Moore, 
Ian O'Dwyer, Gustavo Orellana, 
Stephen Osborne, Heather Owen, Stephen Padin,
Felipe Pedreros,
Ashley Perko, Alan Robinson,
Jacklyn Sanders, Dale Sanford, Yunior Savon, 
Mary Soria, Alex Sugarbaker, 
David Sutton,
Matias Vidal, Liza Volkova, 
Edward Wollack, 
Stephanie Xenos, Octavio Zapata, Mark Zaskowski and Joseph Zuntz.
\end{acknowledgements}

\begin{deluxetable*}{cc|cc|cc|c}
\tablecaption{QUIET Polarization Power Spectra\label{tab:spectra2}}
\tablecolumns{9}
\tablehead{\multicolumn{2}{c}{} & \multicolumn{2}{c}{ML Pipeline} & \multicolumn{2}{c}{PCL Pipeline} & Syst. }
\tablecomments{Tabulated values are given in CMB thermodynamic units
  of $\mu\textrm{K}^2$, scaled as $C_{\ell}\ell(\ell+1)/2\pi$. 
 We present the results from both the ML and PCL pipelines; they are in
 excellent statistical agreement.
 The
  column $EE$/$q$ shows the $EE$ power spectrum normalized 
 to $q=1$, as plotted in Figure~\ref{fig:cl_wband_scaled}.
 The fit value of $q$ is also shown in the table, where the first and second
 errors are statistical and systematic, respectively.
 The column of $BB$--power-spectrum values also provides 95\% confidence level upper limits in parentheses.
 We also list two relevant systematic-error contributions besides the
 uncertainty from the responsivity calibration: the fractional error due to
 the uncertainty of the beam window function, and the total systematic
 error in the $EB$ power spectrum in units of $\mu\textrm{K}^2$.
 Note that they are both highly correlated among $\ell$ bins.
 We assume the $\Lambda$CDM prediction (i.e., $q=1$) for
 the $C_\ell^{EE}$ spectrum sourcing the systematic error in the $EB$ power;
 the $EB$ systematic error estimate should be multiplied by the fit value of $q$ to 
 directly compare with the presented $EB$ power spectrum.
 }
\startdata
$\ell_{\textrm{min}}$ & $\ell_{\textrm{max}}$ & $EE$  & $EE$/$q$ & $EE$ & $EE$/$q$ & Window Function \\[1pt]
\hline
&&&&&&\\[-6pt]
   26 &  75 &   $\phm0.28^{+0.17}_{-0.14}$ &   $\phm0.21^{+0.13}_{-0.11}$
            &   $\phm0.29^{+0.13}_{-0.10}$ &   $\phm0.24^{+0.11}_{-0.08}$ & 0.00 \\[2pt]
   76 & 125 &   $\phm1.06^{+0.24}_{-0.21}$ &   $\phm0.78^{+0.18}_{-0.16}$
            &   $\phm0.82^{+0.20}_{-0.17}$ &   $\phm0.67^{+0.16}_{-0.14}$ & 0.01 \\[2pt]
  126 & 175 &   $\phm1.80^{+0.35}_{-0.32}$ &   $\phm1.34^{+0.26}_{-0.23}$
            &   $\phm1.35^{+0.30}_{-0.27}$ &   $\phm1.11^{+0.25}_{-0.22}$ & 0.01 \\[2pt]
  176 & 225 &   $\phm0.93^{+0.37}_{-0.33}$ &   $\phm0.69^{+0.28}_{-0.25}$
            &   $\phm1.48^{+0.41}_{-0.37}$ &   $\phm1.21^{+0.34}_{-0.30}$ & 0.02 \\[2pt]
  226 & 275 &  \phn$\phm3.5^{+0.7}_{-0.6}$ &  \phn$\phm2.6^{+0.5}_{-0.4}$
            &  \phn$\phm2.9^{+0.6}_{-0.5}$ &  \phn$\phm2.3^{+0.5}_{-0.5}$ & 0.02 \\[2pt]
  276 & 325 &     $\phm13.5^{+1.3}_{-1.3}$ &     $\phm10.0^{+1.0}_{-1.0}$
            &     $\phm12.7^{+1.3}_{-1.2}$ &     $\phm10.4^{+1.1}_{-1.0}$ & 0.03 \\[2pt]
  326 & 375 &     $\phm24.8^{+2.0}_{-2.1}$ &     $\phm18.3^{+1.5}_{-1.5}$
            &     $\phm21.1^{+1.8}_{-1.7}$ &     $\phm17.3^{+1.5}_{-1.4}$ & 0.04 \\[2pt]
  376 & 425 &     $\phm30.6^{+2.6}_{-2.4}$ &     $\phm22.7^{+1.9}_{-1.8}$
            &     $\phm27.2^{+2.2}_{-2.1}$ &     $\phm22.3^{+1.8}_{-1.7}$ & 0.05 \\[2pt]
  426 & 475 &     $\phm19.8^{+2.3}_{-2.4}$ &     $\phm14.7^{+1.7}_{-1.8}$
            &     $\phm18.3^{+2.2}_{-2.1}$ &     $\phm15.0^{+1.8}_{-1.7}$ & 0.05 \\[2pt]
  476 & 525 &     $\phm15.5^{+2.7}_{-2.6}$ &     $\phm11.5^{+2.0}_{-1.9}$
            &  \phn$\phm9.1^{+2.3}_{-2.1}$ &  \phn$\phm7.4^{+1.9}_{-1.7}$ & 0.06 \\[2pt]
  526 & 575 &     $\phm14.1^{+3.3}_{-3.1}$ &     $\phm10.5^{+2.4}_{-2.3}$
            &     $\phm11.4^{+3.0}_{-2.8}$ &  \phn$\phm9.3^{+2.5}_{-2.3}$ & 0.07 \\[2pt]
  576 & 625 &     $\phm24.7^{+4.5}_{-4.3}$ &     $\phm18.3^{+3.3}_{-3.2}$
            &     $\phm19.5^{+4.2}_{-4.0}$ &     $\phm16.0^{+3.5}_{-3.3}$ & 0.08 \\[2pt]
  626 & 675 &       $\phm49^{+7}_{-7}$\phe &       $\phm36^{+5}_{-5}$\phe
            &       $\phm40^{+6}_{-6}$\phe &       $\phm33^{+5}_{-5}$\phe & 0.09 \\[2pt]
  676 & 725 &       $\phm40^{+8}_{-8}$\phe &       $\phm30^{+6}_{-6}$\phe
            &       $\phm37^{+7}_{-7}$\phe &       $\phm31^{+6}_{-5}$\phe & 0.10 \\[2pt]
  726 & 775 &         $\phm25^{+10}_{-10}$ &       $\phm18^{+7}_{-7}$\phe
            &       $\phm21^{+8}_{-8}$\phe &       $\phm17^{+7}_{-7}$\phe & 0.11 \\[2pt]
  776 & 825 &         $\phm20^{+14}_{-13}$ &         $\phm15^{+10}_{-10}$
            &         $\phm30^{+12}_{-11}$ &          $\phm25^{+10}_{-9}$ & 0.12 \\[2pt]
  826 & 875 &         \phn$-9^{+17}_{-16}$ &         \phn$-7^{+13}_{-12}$
            &         \phn$-8^{+14}_{-13}$ &         \phn$-7^{+11}_{-11}$ & 0.14 \\[2pt]
  876 & 925 &         $\phm93^{+29}_{-28}$ &         $\phm69^{+21}_{-21}$
            &         $\phm68^{+24}_{-22}$ &         $\phm55^{+19}_{-18}$ & 0.15 \\[2pt]
  926 & 975 &         $\phm35^{+39}_{-37}$ &         $\phm26^{+29}_{-27}$
            &         $\phm85^{+32}_{-30}$ &         $\phm69^{+26}_{-25}$ & 0.17 \\[2pt]
      &
      &  \multicolumn{2}{c}{$q = 1.35 \pm 0.05  {}^{+0.26}_{-0.22}$}
      &  \multicolumn{2}{c}{$q = 1.22 \pm 0.04  {}^{+0.22}_{-0.17}$} \\[2pt]
\hline
\\[-4pt]
$\ell_{\textrm{min}}$ & $\ell_{\textrm{max}}$ & $BB$ (95\% UL) & $EB$ & $BB$ (95\% UL) & $EB$ & $EB$ syst. \\[1pt]
\hline
&&&&&&\\[-6pt]
   26 &  75 &      $-0.08^{+0.10}_{-0.08}$     ($0.19$) &   $\phm0.07^{+0.03}_{-0.04}$
            &   $\phm0.03^{+0.07}_{-0.05}$     ($0.18$) &      $-0.04^{+0.04}_{-0.04}$ & $\pm 0.01$ \\[2pt]
   76 & 125 &   $\phm0.26^{+0.15}_{-0.14}$     ($0.55$) &      $-0.04^{+0.11}_{-0.11}$
            &   $\phm0.24^{+0.15}_{-0.12}$     ($0.50$) &   $\phm0.03^{+0.11}_{-0.11}$ & $\pm 0.02$ \\[2pt]
  126 & 175 &      $-0.25^{+0.16}_{-0.14}$     ($0.23$) &   $\phm0.12^{+0.16}_{-0.16}$
            &      $-0.22^{+0.15}_{-0.13}$     ($0.23$) &   $\phm0.17^{+0.15}_{-0.14}$ & $\pm 0.03$ \\[2pt]
  176 & 225 &   $\phm0.53^{+0.31}_{-0.28}$     ($1.09$) &   $\phm0.13^{+0.22}_{-0.22}$
            &   $\phm0.39^{+0.32}_{-0.28}$     ($0.95$) &   $\phm0.17^{+0.24}_{-0.23}$ & $\pm 0.02$ \\[2pt]
  226 & 275 &      $-0.59^{+0.37}_{-0.34}$     ($0.52$) &   $\phm0.11^{+0.33}_{-0.34}$
            &      $-0.60^{+0.35}_{-0.31}$     ($0.49$) &   $\phm0.12^{+0.33}_{-0.32}$ & $\pm 0.08$ \\[2pt]
  276 & 325 &     \phn$-0.4^{+0.6}_{-0.5}$  \phn($1.0$) &     \phn$-0.1^{+0.6}_{-0.6}$
            &     \phn$-0.2^{+0.5}_{-0.5}$  \phn($1.0$) &  \phn$\phm0.1^{+0.6}_{-0.6}$ & $\pm 0.29$ \\[2pt]
  326 & 375 &     \phn$-0.5^{+0.7}_{-0.7}$  \phn($1.2$) &  \phn$\phm2.3^{+0.8}_{-0.9}$
            &     \phn$-0.1^{+0.8}_{-0.7}$  \phn($1.5$) &  \phn$\phm2.0^{+0.9}_{-0.8}$ & $\pm 0.59$ \\[2pt]
  376 & 425 &     \phn$-0.1^{+1.0}_{-1.0}$  \phn($2.1$) &  \phn$\phm1.2^{+1.1}_{-1.1}$
            &  \phn$\phm0.4^{+1.1}_{-1.0}$  \phn($2.5$) &  \phn$\phm0.9^{+1.1}_{-1.1}$ & $\pm 0.75$ \\[2pt]
  426 & 475 &  \phn$\phm1.6^{+1.5}_{-1.4}$  \phn($4.4$) &  \phn$\phm1.8^{+1.3}_{-1.3}$
            &  \phn$\phm1.2^{+1.4}_{-1.3}$  \phn($3.7$) &  \phn$\phm2.5^{+1.2}_{-1.2}$ & $\pm 0.55$ \\[2pt]
  476 & 525 &     \phn$-0.1^{+1.9}_{-1.8}$  \phn($4.0$) &  \phn$\phm0.9^{+1.6}_{-1.6}$
            &     \phn$-2.5^{+1.7}_{-1.6}$  \phn($2.3$) &  \phn$\phm0.2^{+1.4}_{-1.3}$ & $\pm 0.30$ \\[2pt]
  526 & 575 &  \phn$\phm4.2^{+2.8}_{-2.7}$  \phn($9.2$) &  \phn$\phm0.3^{+2.0}_{-2.0}$
            &  \phn$\phm2.3^{+2.4}_{-2.3}$  \phn($6.6$) &     \phn$-1.6^{+1.9}_{-2.0}$ & $\pm 0.28$ \\[2pt]
  576 & 625 &  \phn$\phm3.8^{+3.7}_{-3.5}$     ($10.5$) &  \phn$\phm3.4^{+2.7}_{-2.8}$
            &     \phn$-1.1^{+3.1}_{-2.9}$  \phn($5.6$) &  \phn$\phm1.8^{+2.6}_{-2.5}$ & $\pm 0.70$ \\[2pt]
  626 & 675 &  \phn$\phm9.2^{+5.2}_{-4.9}$     ($18.3$) &  \phn$\phm7.8^{+4.1}_{-4.0}$
            &  \phn$\phm6.3^{+4.4}_{-4.1}$     ($13.8$) &  \phn$\phm4.8^{+3.5}_{-3.5}$ & $\pm 1.25$ \\[2pt]
  676 & 725 &    \phn$\phm3^{+7}_{-6}$\phe \phn\phn\phn($16$) &    \phn$\phm1^{+5}_{-5}$\phe
            &    \phn$\phm8^{+6}_{-6}$\phe \phn\phn\phn($18$) &  \phn$\phm6.7^{+4.8}_{-4.8}$ & $\pm 1.40$ \\[2pt]
  726 & 775 &       \phn$-8^{+9}_{-8}$\phe \phn\phn\phn($13$) &    \phn$\phm3^{+6}_{-6}$\phe
            &    \phn$\phm1^{+8}_{-8}$\phe \phn\phn\phn($17$) &          $-11^{+5}_{-6}$\phe & $\pm 1.06$ \\[2pt]
  776 & 825 &      \phn$\phm7^{+13}_{-12}$ \phn\phn\phn($31$) &          $-19^{+9}_{-9}$\phe
            &      \phn$\phm5^{+11}_{-10}$ \phn\phn\phn($26$) &          $-14^{+8}_{-8}$\phe & $\pm 0.68$ \\[2pt]
  826 & 875 &         $\phm10^{+18}_{-17}$ \phn\phn\phn($44$) &         $\phm15^{+12}_{-12}$
            &      \phn$\phm2^{+15}_{-14}$ \phn\phn\phn($31$) &         \phn$-9^{+10}_{-10}$ & $\pm 0.55$ \\[2pt]
  876 & 925 &         $\phm46^{+27}_{-26}$ \phn\phn\phn($93$) &      \phn$\phm4^{+19}_{-19}$
            &         \phn$-3^{+19}_{-18}$ \phn\phn\phn($37$) &      \phn$\phm0^{+15}_{-15}$ & $\pm 0.93$ \\[2pt]
  926 & 975 &            $-52^{+35}_{-33}$ \phn\phn\phn($44$) &      \phn$\phm0^{+25}_{-25}$
            &            $-28^{+27}_{-26}$ \phn\phn\phn($41$) &         $\phm17^{+19}_{-19}$ & $\pm 1.59$
\enddata
\end{deluxetable*}

\end{document}